\documentclass[aps,prb,twocolumn,superscriptaddress,groupedaddress]{revtex4-2}  % for review and submission

\usepackage{amsmath}
\usepackage{graphicx}
\usepackage{subfigure}
\usepackage{color}
\usepackage{amsfonts}
\usepackage{tikz}
\usepackage{esint}
\usepackage{mathrsfs}
\usepackage{hyperref}
\hypersetup{
     colorlinks = true,
     citecolor  = blue,  % cite
     linkcolor  = blue, % ref
     urlcolor = blue,
     filecolor = blue
}
\begin{document}

\title{Evidence of triplet superconductivity in Bi/Ni bilayers: Theoretical analysis of point contact Andreev reflection spectroscopy results}

\author{Jia-Cheng He}
\email{jche14@fudan.edu.cn}

\author{Yan Chen}
%\thanks{Corresponding author}
\email{yanchen99@fudan.edu.cn}
\affiliation{Department of Physics and State Key Laboratory of Surface Physics, Fudan University, Shanghai 200433, China}
\date{\today}

\begin{abstract}
A theoretical formalism of Andreev reflection is employed to provide theoretical support for distinguishing between the singlet pairing and the triplet pairing by the point contact Andreev reflection (PCAR) experiments. We utilize our theoretical curves to fit the data of the PCAR experiment on unconventional superconductivity in the Bi/Ni bilayer [\href{https://arxiv.org/abs/1810.10403}{arXiv:1810.10403}] and find the Anderson-Brinkman-Morel (ABM) state satisfies the main characteristics of the experimental data. The chiral cross-section of the ABM state might explain well the broken time-reversal symmetry determined by the polar Kerr effect measurements and the time-domain THz spectroscopy in Bi/Ni bilayers. Moreover, the Andreev reflection spectra of the Balian-Werthamer state and the chiral $p$-wave state are also presented.

\end{abstract}

\maketitle

\section{introduction}
At the interface of a normal metal (N) and a superconductor (S), incident electrons from the N side are reflected as holes and transmitted into the S side as Cooper pairs. This is the Andreev reflection (AR) \cite{Andreev} process which is the primary mechanism of electron transport across an N-S interface. The AR spectra can be used to study the characteristics of the superconducting gap, including its symmetry and magnitude. As the quantitative model for the AR process, the Blonder-Tinkham-Klapwijk (BTK) model \cite{BTK} has been used to study the isotropic gap of the BCS superconductor. The model proposed by Kashiwaya and Tanaka \cite{Kashiwaya,Tanaka} can be used to describe the anisotropic gap superconductor by analysis of the corresponding conductance spectra of the N-S junction. The Mazin model \cite{Mazin} can be used to analyze the fully polarized current across an N-S junction quantitatively. The unified model \cite{ChenTesanovic} is valid for the quantitative analysis of the current with arbitrary polarization.

As the well-known triplet pairing, the $p$-wave state was firstly found in the electrically neutral superfluid $^{3}$He \cite{Osheroff}. However, the $p$-wave state has never been verified in solids experimentally. The triplet pairing superconductors are often associated with topological superconductivity \cite{Mizushima} and applications in quantum computing and spintronics \cite{Nayak,Romeo}. There are some candidates for triplet superconductivity, including the heavy-fermion superconductors (e.g., UPt$_{3}$) \cite{Stewart,Ishida,Saxena,Aoki,Huy}, the superconductors with broken inversion symmetry \cite{Nishiyama,Bauer}, and the well-known Sr$_{2}$RuO$_{4}$ \cite{Mackenzie,Maeno,Liu}. Recently the AR spectroscopy of Bi/Ni bilayers \cite{Jin} might indicate the existence of $p$-wave superconductivity in solids. Moreover, the Bi/Ni bilayer itself is interesting enough. The single-crystal Bi (110), observed bulk superconductivity below 0.53 mK \cite{Prakash52}, is epitaxially grown on the weak ferromagnetic Ni (100) layer to yield a Bi/Ni bilayer whose superconducting transition temperature is enhanced to 4 K \cite{Jin}. Therefore, this bilayer system has attracted much research interest \cite{Jin,ZGH,PhysRevB.99.064504,PhysRevB.101.174514,Gonge1602579,NodelessBiNiPhysRevLett.122.017002}.

In this paper, we employ a theoretical formalism of AR by the four-component wave function to naturally obtain the singlet pairing case or the triplet pairing case. Our work provides theoretical support for point contact Andreev reflection (PCAR) experiments to distinguish between the singlet pairing and the triplet pairing. The AR conductance of singlet pairing superconductivity depends on spin polarization, but the case of triplet pairing superconductivity is not related to spin polarization.

This paper is organized as follows. In Section \ref{Sec:conducanceformula}, our formalism is introduced briefly. In Section \ref{Sec:conductanceABM2D}, we follow the procedure of the formalism to obtain the results of the Anderson-Brinkman-Morel (ABM) state. In Section \ref{sec:fittingandanalysis}, by utilizing the theoretical curves to fit the data of the PCAR experiment on the unusual superconductivity of the epitaxial Bi/Ni bilayer, we find that the theoretical conductance spectra of the ABM state can describe well the main features of the experimental conductance spectra. In Section \ref{sec:ABM_arbitrary_cross-section},  we discuss the AR for an arbitrary cross-section of the 3D gap of the ABM state. The chiral cross-section of the gap of the ABM state might explain well the polar Kerr effect measurements and the time-domain THz spectroscopy in the Bi/Ni bilayer. The discussion and conclusion will be given in Section \ref{Sec:DiscussionConclusion}.

\section{conductance formula}\label{Sec:conducanceformula}
The plane wave at the normal metal side of the N-S junction can be expressed by using the four-component wave function
\begin{equation}\label{eq:normalstate}
\psi_{N}(\mathbf{ r})=e^{i\mathbf{ k}_{\parallel}\cdot \mathbf{ r}_{\parallel}}\left(\begin{array}{c}
e^{ik_{+}x}+be^{-ik_{+}x}\\
0\\
a_{2}e^{ik_{-}x}\\
a_{1}e^{(\alpha+i)k_{-}x}
\end{array}\right),
\end{equation}
where $\mathbf{k}_{\parallel}$ and $k_{+}$ ($k_{-}$) are wave vector components parallel and vertical to the N-S junction interface, respectively. The subscripts $+$ and $-$ denote the electron-like and hole-like quasiparticles, respectively. The dimensionless real number $\alpha$ is related to spin polarization $P$ \cite{ChenTesanovic}, i.e., $P=\frac{\alpha^{2}}{4+\alpha^{2}}$. $e^{ik_{+}x}$ is a spin-up incident plane wave. $be^{-ik_{+}x}$ is a normal reflection wave. $a_{1}e^{(\alpha+i)k_{-}x}$ is a normal AR wave which is evanescent and spin up. $a_{2}e^{ik_{-}x}$ is an unconventional AR wave with spin down. Since we can always choose the spin direction of the incident electron as the positive $s_{Z}$ direction, the second row, which represents the electron with spin down, is $0$. As a result, Eq. (\ref{eq:normalstate}) has no loss of generality. In addition, the coefficients ($a_{1}$, $a_{2}$ and $b$) can be calculated using the boundary conditions given below.

The wave function at the superconductor side is given by
\begin{equation}\label{eq:superstate}
\psi_{S}(\mathbf{ r})=e^{i\mathbf{ k}_{\parallel}\cdot \mathbf{ r}_{\parallel}}\bigg\{ce^{iq_{+}x}\left(\begin{array}{c}
u^{(+)}_{\mathbf{k}\uparrow\uparrow}\\
u^{(+)}_{\mathbf{k}\downarrow\uparrow}\\
v^{*(+)}_{\mathbf{-k}\uparrow\uparrow}\\
v^{*(+)}_{\mathbf{-k}\downarrow\uparrow}
\end{array}\right)+de^{-iq_{-}x}\left(\begin{array}{c}
u^{(-)}_{\mathbf{k}\uparrow\uparrow}\\
u^{(-)}_{\mathbf{k}\downarrow\uparrow}\\
v^{*(-)}_{\mathbf{-k}\uparrow\uparrow}\\
v^{*(-)}_{\mathbf{-k}\downarrow\uparrow}
\end{array}\right)\bigg\},
\end{equation}
where the superscripts $(+)$ and $(-)$ denote electron-like and hole-like quasiparticles, respectively. $u_{\mathbf{k}ss^{\prime}}$ and $v_{\mathbf{k}ss^{\prime}}$ are coherence factors originating from the equation \cite{Sigrist}
\begin{equation}\label{eq:quasiparticle}
a_{\mathbf{k}s}=\sum_{s'}\left(u_{\mathbf{k}ss'}\alpha_{\mathbf{k}s'}+v_{\mathbf{k}ss'}\alpha_{-\mathbf{k}s'}^{\dagger}\right),
\end{equation}
where $s$ is the spin index. $a_{\mathbf{k}s}$ is the electron's annihilate operator, and $\alpha_{\mathbf{k}s'}$ is the quasiparticle's annihilate operator. In this paper, we only consider the unitary solution of the superconductor. More details about Eq. (\ref{eq:superstate}) can be found in Appendix \ref{sec:appendixunitarysolution}. For simplification, we take $q_{\pm}\approx k_{\pm}\approx k_{F}\cos\theta_{N}$ where $k_{F}$ is the Fermi wave vector size at the normal metal side, and $\theta_{N}$ is the angle between the direction of the incident electron and the normal to the interface. 

The boundary conditions are given by
\begin{flalign}
\psi(x=0^{+})-\psi(x=0^{-})&=0,
              \nonumber\\
\frac{\hbar^{2}}{2m}\psi'_{x}\bigg|_{x=0^{-}}-\frac{\hbar^{2}}{2m}\psi'_{x}\bigg|_{x=0^{+}}+U\psi\bigg|_{x=0}&=0.
\label{eq:boundary}
\end{flalign}
For the ballistic transport, we utilize the 2D BTK model \cite{Kashiwaya,Tanaka,Daghero_2010,Daghero_2011} and thus the normalized conductance with a bias voltage $V$ is
\begin{equation}\label{eq:generalconductance}
\sigma(eV)=\frac{\int d\mathbf{k}_{\parallel}g^{T}(eV)}{\int d\mathbf{k}_{\parallel}g^{T}(\infty)},
\end{equation}
where\quad$g^{T}(eV)=\int_{0}^{1}g\left(|eV+\frac{1}{\beta}\ln \frac{1-f}{f}|\right)df$,\quad$g(E)=1+|a_{1}(E)|^{2}+|a_{2}(E)|^{2}-|b(E)|^{2}$ and $\beta=1/k_{B}T$. $k_{B}$ is the Boltzmann constant. $g(\infty)=\frac{1}{(Z/\cos\theta_{N})^{2}+1}$ with $Z=mU/\hbar^{2}k_{F}$ is the transparency of the N-S interface and is a function of the incident angle $\theta_{N}$ of the electron \cite{Daghero_2010}. To make it more concrete, the normalized conductance formula is rewritten as
\begin{equation}\label{eq:normalizedcd}
\sigma(eV)=\frac{\int_{-\pi/2}^{\pi/2}g^{T}(eV)\cos\theta_{N} d\theta_{N}}{\int_{-\pi/2}^{\pi/2}g^{T}(\infty)\cos\theta_{N} d\theta_{N}}.
\end{equation}

The 2D model for AR cannot fully describe the case of the 3D anisotropic gap and is only applicable to the case of the isotropic gap, such as the $s$-wave case, and a specific cross-section of the 3D anisotropic gap. The 3D model for AR can be applied to any $\mathbf{k}$-dependent 3D order parameter and any shape of the Fermi surface \cite{Daghero_2010, Daghero_2011}, and it has been successfully used to fit spectra in a variety of compounds \cite{GONNELLI201372,doi:10.1063/1.4794994, Gonnelli:2016vu}. We will present the method by the ABM state case for an arbitrary cross-section of the 3D anisotropic gap.

Following the steps described above, one can obtain expressions for the coefficients $a_{1}$, $a_{2}$, $b$, $c$, and $d$ of the $s$-wave and $d$-wave cases, which are consistent with Ref. \cite{ChenTesanovic} and Ref. \cite{Kashiwaya}, respectively. The conductance of the $s$-wave or $d$-wave case depends on spin polarization, which is the significant characteristic of the singlet pairing case; that is, the conductance peak of the singlet pairing case disappears, and the conductance within the gap voltage approaches zero as spin polarization increases \cite{ChenTesanovic}.

Note that we have presented the part of our theoretical formalism in the supplementary materials of Ref. \cite{ZGH}, and our theoretical fitting curves to the experimental data can also be found in Ref. \cite{ZGH}. This paper demonstrates a more detailed version of our theoretical formalism. Furthermore, the following sections contain some in-depth discussions of the theoretical fitting curves to the experimental data.

\begin{figure}
\centering
\includegraphics[width=0.48\textwidth]{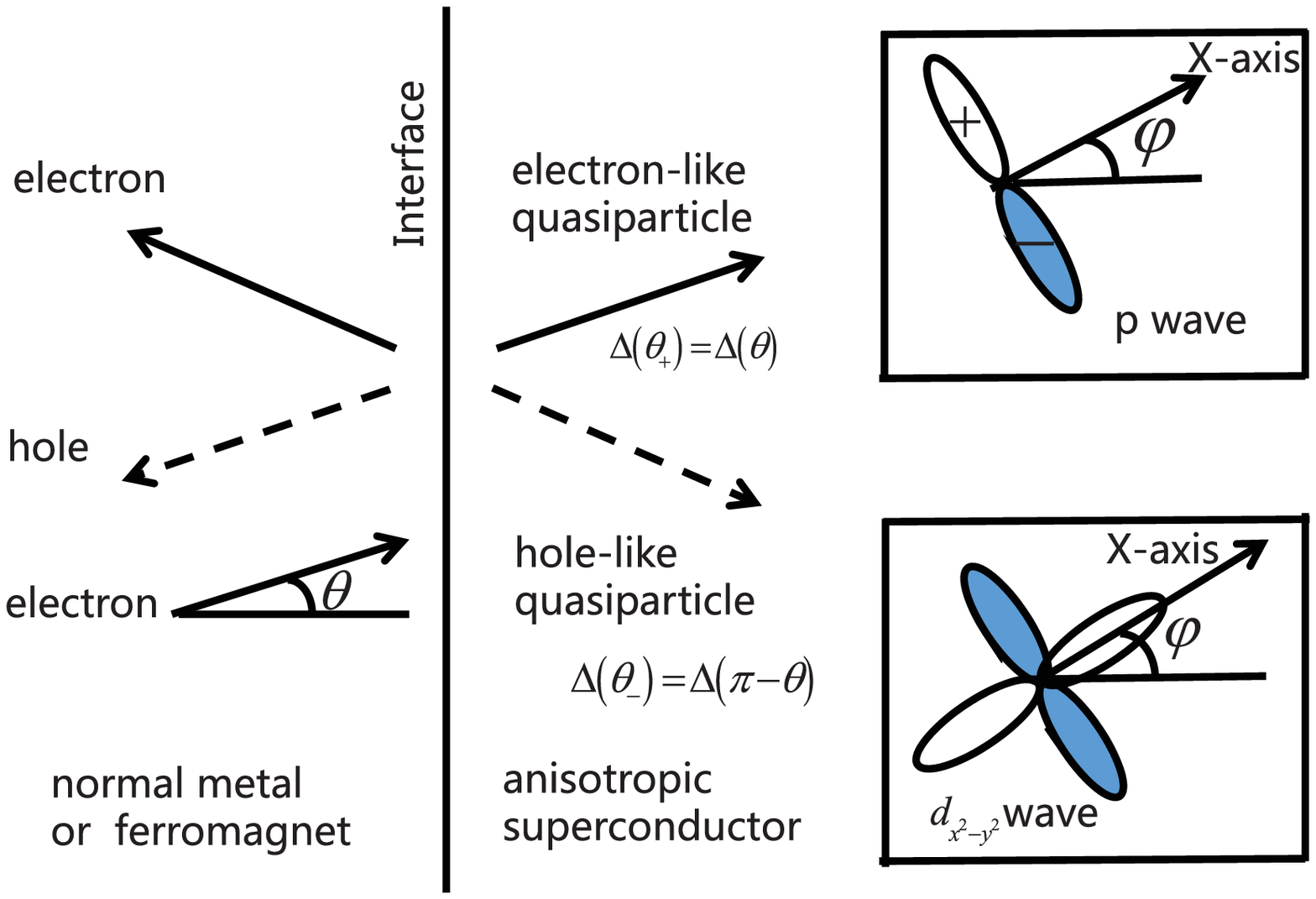}
\caption{\label{fig:dandpwavedraft}(Color online) The process of transmission and reflection at the N-S junction. The parameter $\theta$ represents the incident angle of the electron. The parameter $\varphi$ represents the angle between the $x$-axis of the $p$-wave ($d$-wave) and the normal direction of the interface. }
\end{figure}

\section{conductance of the ABM state}\label{Sec:conductanceABM2D}
There are two important  $p$-wave states in the triplet pairing case \cite{Sigrist,Vollhardt}. The first state is the ABM state \cite{Anderson,Brinkman,Leggett}, and its superconducting order parameter has the form:
\begin{equation}\label{eq:ABMorderparameter}
\hat{\Delta}(\mathbf{ k})=\Delta\left(\begin{array}{cc}
-e^{i\phi_{\mathbf{k}}}\sin\theta_{\mathbf{k}}&0\\
0&e^{i\phi_{\mathbf{k}}}\sin\theta_{\mathbf{k}}
\end{array}\right).
\end{equation}
Here $\theta_{\mathbf{k}}$ and $\phi_{\mathbf{k}}$ are related to the spatial direction $(\theta_{\mathbf{k}}, \phi_{\mathbf{k}})$ on the Fermi surface. More details about the ABM state can be found in Appendix \ref{sec:propertiesofABMorder}. The relation $\hat{\Delta}(\mathbf{ k})\hat{\Delta}^{\dagger}(\mathbf{ k})=\Delta^{2}\sin^{2}\theta_{\mathbf{k}}\hat{\sigma}_{0}$ indicates that the ABM state belongs to the unitary solution. Thus its coherence factors are elements of matrixes $\hat{u}_{\mathbf{k}}=\sqrt{\frac{1}{2}(1+\epsilon(\mathbf{ k})/E_{\mathbf{k}})}\hat{\sigma}_{0}$, and $\hat{v}_{\mathbf{k}}=\Delta e^{i\phi_{\mathbf{k}}}\sin\theta_{\mathbf{k}}\hat{\sigma}_{z}/\sqrt{2E_{\mathbf{k}}(E_{\mathbf{k}}+\epsilon(\mathbf{ k}))}$, ($u_{\downarrow\uparrow}^{(\pm)}=v_{\downarrow\uparrow}^{(\pm)}=0)$.

We choose the coordinate system $I$ in which the $I_{x}-I_{y}$ plane lies in the N-S interface and the $I_{z}$-axis points to the S side. We restrict the axis of symmetry of the ABM state's gap to the $I_{x}-I_{z}$ plane, so the angle part of the axis of symmetry is $(\theta_{n},\varphi_{n})\equiv(\omega,0)$. For instance, the form of the ABM state in Eq. (\ref{eq:ABMorderparameter}) corresponds to the case of $\omega=0$. The electron across the N-S interface has four trajectories \cite{BTK}: electron reflection, hole reflection, electron-like quasiparticle transmission, and hole-like quasiparticle transmission, as shown in Fig. \ref{fig:dandpwavedraft}. These four trajectories lie in the same plane according to the translation invariance in the interface \cite{Kashiwaya}. We use $(\theta_{\mathbf{k}_{1}}, \phi_{\mathbf{k}_{1}})$ to denote the direction of the wave vector $\mathbf{k}_{1}$ of the incident electron. Therefore, $\phi_{\mathbf{k}_{1}}$ and $\phi_{\mathbf{k}_{1}}+\pi$ can be used to determine the plane where the AR process occurs, and the parameter $\phi_{0}$ given by $\phi_{0}=\phi_{\mathbf{k}_{1}}$ is enough to denote this plane. The parameters $\omega$ and $\phi_{0}$ together determine the configuration of connecting the tip and the sample of the ABM state in the PCAR experiment.

The ABM state's gap is anisotropic in three dimensions, but we first study the gap's cross-section in the case of $\phi_{0}=0$ to get a sense of it. The other parameter $\omega$ of this cross-section is free, and the corresponding form of the order parameter is given by $\mp\Delta\sin(\theta_{\mathbf{k}}\mp\omega)\hat{\sigma}_{z}$ with $\theta_{\mathbf{k}},\omega\in\left[0,\pi\right]$. We apply the polar coordinates to this cross-section, with the polar angle measured from the positive $I_{z}$-axis of the $I_{x}-I_{z}$ plane, for simplification. We use $\theta_{\mathbf{q}}$ to denote this polar angle, with the vector $\mathbf{q}$ in this cross-section to avoid confusion. Therefore, the ABM state's order parameter projected onto this cross-section can be written as $\Delta\sin(\theta_{\mathbf{q}}-\omega)\hat{\sigma}_{z}$ with $\theta_{\mathbf{q}}\in\left[0,2\pi\right]$. The calculated coefficients $a_{1}$, $a_{2}$, $b$, $c$, and $d$ in this cross-section case are listed in Table \ref{tab:ABMresult} of Appendix \ref{sec:propertiesofABMorder}, where $Z$ is given by $Z=mU/\hbar^{2}k_{F}$ and $\Gamma$ is the inelastic scattering factor \cite{Dynes}. The corresponding process of AR is shown in Fig. \ref{fig:dandpwavedraft}. Other parameters can be illustrated in Fig. \ref{fig:dandpwavedraft}. The transmitted electron-like quasiparticle and the hole-like quasiparticle have different effective pair potentials $\Delta(\theta_{+})$ and $\Delta(\theta_{-})$, with $\theta_{+}=\theta$ and $\theta_{-}=\pi-\theta$, respectively. We can easily understand from $a_{1}\equiv0$ that these coefficients are not related to the parameter $\alpha$, according to Eq. (\ref{eq:normalstate}). Thus these coefficients are independent of spin polarization $P$, according to $P=\frac{\alpha^{2}}{4+\alpha^{2}}$. Therefore, the conductance of the ABM state is not related to spin polarization, contrary to the singlet pairing case.

\begin{figure}
\centering
\includegraphics[width=0.48\textwidth]{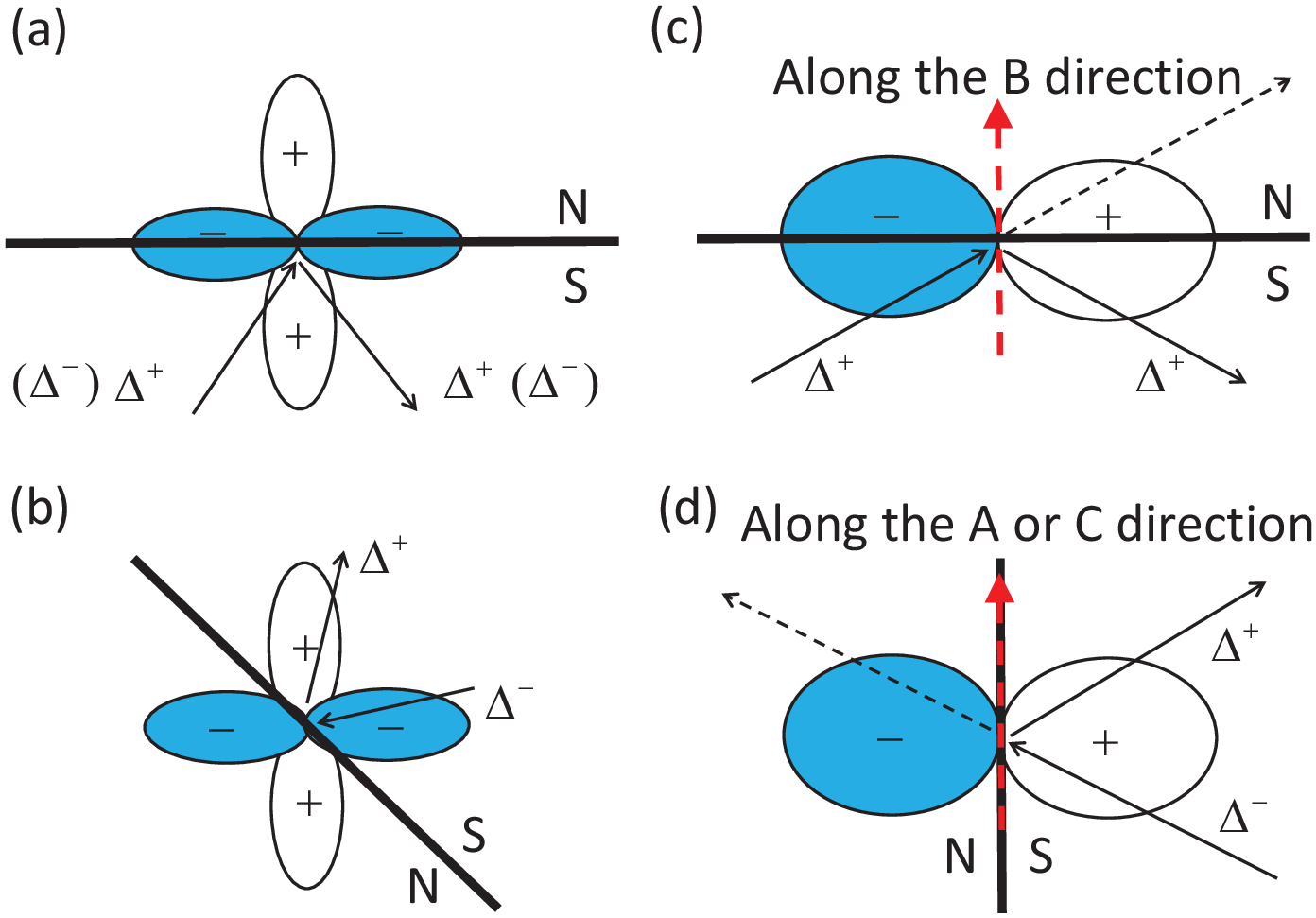}
\caption{\label{fig:pdtunnels}(Color online) The incident quasiparticle from the S side will change its phase of the pair potential after being reflected by the N-S interface in different cases. The thick dark line represents the N-S interface. The red dotted line with an arrow represents the axis of symmetry of the gap of the ABM state. (a) The $d$-wave case of $\varphi=0$ ($\varphi$ is explained in Fig. \ref{fig:dandpwavedraft}), (b) the $d$-wave case of $\varphi=\pi/4$, (c) the cross-section of the 3D gap of the ABM state in the case of $\omega=0$ and $\phi_{0}=0$, (d) the cross-section of the 3D gap of the ABM state in the case of $\omega=\pi/2$ and $\phi_{0}=0$. The zero-bias conductance peak of the PCAR spectra originates from the $\pi$ phase difference of the pair potential of the reflected quasiparticle to that of the incident quasiparticle.  }
\end{figure}

\begin{figure}
\centering
\includegraphics[width=0.48\textwidth]{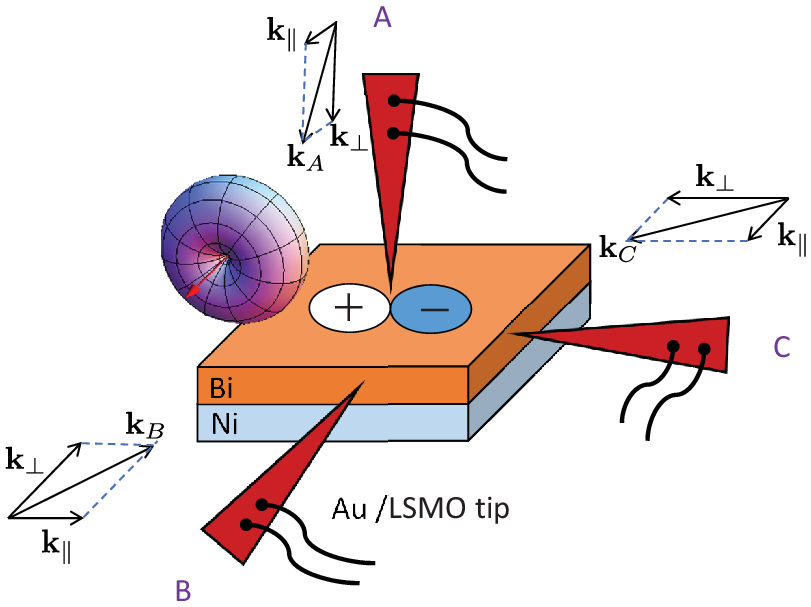}
\caption{\label{fig:3Ddemonstrationmini}(Color online) Schematic illustration of the PCAR experimental work on Bi/Ni bilayers \cite{Jin,ZGH}. They measured the conductance spectra of epitaxial Bi/Ni bilayers (based on various thicknesses of the Ni layer (0-7.5 nm) and the Bi layer (0-500 nm) \cite{Jin,ZGH}) in three almost mutually perpendicular directions (i.e., the A, B and C directions). Compared with their experimental conductance spectra, our Figs. \ref{fig:ABMstateangle}(a) and \ref{fig:ABMstateangle}(b) correspond to the A direction case, Figs. \ref{fig:ABMstateangle}(c) and \ref{fig:ABMstateangle}(d) correspond to the B direction case, and Figs. \ref{fig:ABMstateangle}(e) and \ref{fig:ABMstateangle}(f) correspond to the C direction case. Therefore, the ABM state might be indicated in the bulk of the Bi layer, and the axis of symmetry of the gap of the ABM state might be almost parallel to the B direction. }
\end{figure}

\section{The fit and analysis of the experimental data}\label{sec:fittingandanalysis}
Before discussing the conductance spectra of the ABM state, we provide a physical understanding of the Andreev reflection conductance peak to help you understand the results of the PCAR experiments on the Bi/Ni bilayer. It was proposed that the conductance peak in the $d$-wave case originates from the bound states localized around the N-S interface. These bound states decay into the bulk \cite{Kashiwaya}. The N-S interface can be regarded as the node of pair potential for the quasiparticles. The tunneling electrons flow from N to S via the bound states, similar to the resonant tunneling process. Moreover, the conductance peak forms at the energy levels of the bound states.

For example, in the $d$-wave case of $\varphi=\pi/4$ ($\varphi$ is explained in Fig. \ref{fig:dandpwavedraft}), the incident quasiparticle of the negative pair potential $\Delta^{-}$ from the S side is reflected by the N-S interface, as shown in Fig. \ref{fig:pdtunnels}(b). Subsequently, the phase of the pair potential changes from $\pi$ to 0 at the interface. Then the pair potential at the point of incidence is equal to zero (i.e., $\Delta_{I}=0$) due to the overlap between the negative pair potential of the incident quasiparticle and the positive pair potential of the reflected quasiparticle. Therefore, there is a perfect elastic scattering process for quasiparticles at the interface due to the complete destructive interference of the effective pair potential. The reflectivity of quasiparticles is enhanced by increasing the barrier height $Z$, and the large barrier height limit gives rise to high-density bound states at the interface. There will be a zero-energy peak when the bias voltage makes the Fermi energy of the S side slightly higher than that of the N side (vice versa). However, in the $d$-wave case of $\varphi=0$, the incident quasiparticle of the pair potential $\Delta^{-}$ ($\Delta^{+}$) from the S side will not change the phase after being reflected by the interface, as shown in Fig. \ref{fig:pdtunnels}(a). Then the overlap between the identical phases of the incident quasiparticle and the reflected quasiparticle at the point of incidence results in the finite amplitude of the pair potential (i.e., $\Delta_{I}\neq0$). This contradicts that the pair potential at the N-S interface is zero in the case of the zero bias voltage. Thus incident quasiparticles from the S side are not reflected by the N-S interface in the $d$-wave case of $\varphi=0$ and the zero bias voltage. However, if a positive bias voltage $\Delta/e$ is applied, the Fermi energy difference between the S and N sides will become $\Delta$. The pair potential at the N-S interface becomes $\Delta_{I}=\Delta$ since the hole-like quasiparticle excitation needs at least energy $\Delta$. If a negative bias voltage $-\Delta/e$ is applied, the pair potential at the N-S interface will become $\Delta_{I}=\Delta$ since the electron-like quasiparticle excitation needs at least energy $\Delta$. In the large barrier height limit, the two cases give rise to the high density of bound states at the interface, and thus there is a peak near the bias voltage $\Delta/e$ (or $-\Delta/e$).

The ABM state is analyzed similarly. In the case of the cross-section of $\omega=0$ and $\phi_{0}=0$, the incident quasiparticle of the pair potential $\Delta^{+}$ ($\Delta^{-}$) from the S side will not change the phase after being reflected by the interface, as shown in Fig. \ref{fig:pdtunnels}(c). Thus there is a conductance peak near the bias voltage $\Delta/e$ (or $-\Delta/e$) in the large barrier height limit. In the case of $\omega=\pi/2$ and $\phi_{0}=0$, the pair potential $\Delta^{+}$ ($\Delta^{-}$) of the incident quasiparticle from the S side will change a sign after it is reflected by the interface, as shown in Fig. \ref{fig:pdtunnels}(d). This case yields the zero energy conductance peak in the large barrier height limit.

\begin{figure*}
\centering
\includegraphics[width=1\textwidth]{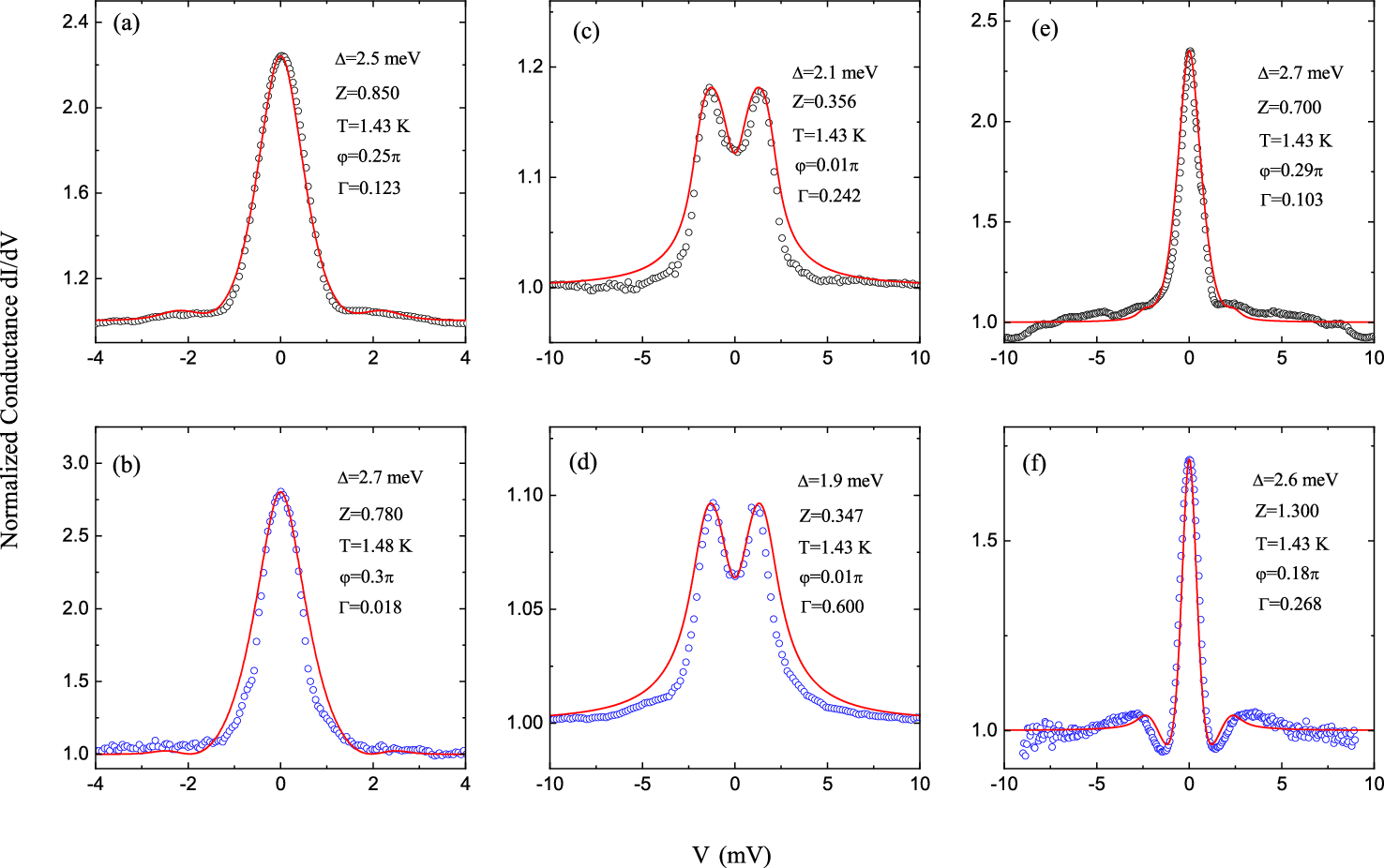}
\caption{\label{fig:ABMstateangle}(Color online) The normalized conductance spectra of the ABM state. The black circles and the blue circles represent the experimental data of the conductance obtained by using the Au tip and the LSMO tip, respectively \cite{ZGH}. The Au tip and the LSMO tip can produce incident electrons with spin unpolarized and highly polarized, respectively. The red lines are theoretical fitting curves based on our theory. Its normalized conductance spectra do not depend on spin polarization, contrary to the singlet pairing case. (a) and (b) show a single peak with values of $\varphi=0.25\pi$ and $\varphi=0.3\pi$, respectively; (c) and (d) show a double peak with a value of $\varphi=0.01\pi$ around zero; (e) and (f) show a single peak with values of $\varphi=0.29\pi$ and $\varphi=0.18\pi$, respectively.}
\end{figure*}

There is the highly unusual superconductivity discovered experimentally in epitaxial Bi/Ni bilayers \cite{Moodera,LeClair}. Evidence of the $p$-wave superconductivity in Bi/Ni bilayers is proposed in Ref. \cite{Jin}. Figure \ref{fig:3Ddemonstrationmini} depicts the PCAR experimental work on Bi/Ni bilayers \cite{ZGH}. The gold (Au) tip and the La$_{0.67}$Sr$_{0.33}$MnO$_{3}$ (LSMO) tip can produce incident electrons with spin unpolarized and highly polarized, respectively. They used these two tips to make vertical contact with the Bi/Ni bilayer surface (i.e., in the A direction). They found that the conductance spectra always show a single peak in the A direction, whether the material of the tip is Au or LSMO. The conductance spectra, in the other two directions parallel to the Bi/Ni bilayer surface, show a double peak in the B direction and a single peak in the C direction, independent of the spin polarization of incident electrons.  

The normalized conductance spectra of the point contacts on Bi/Ni bilayers along the different directions are shown in Fig. \ref{fig:ABMstateangle}. The features of these experimental conductance spectra are completely consistent with our theoretical conductance spectra. In Fig. \ref{fig:ABMstateangle}, the black circles and the blue circles represent the experimental data of the conductance obtained by using the Au tip and the LSMO tip, respectively. The data of Figs. \ref{fig:ABMstateangle}(a) and \ref{fig:ABMstateangle}(b) are measured in the A direction, the data of Figs. \ref{fig:ABMstateangle}(c) and \ref{fig:ABMstateangle}(d) are measured in the B direction, and the data of Figs. \ref{fig:ABMstateangle}(e) and \ref{fig:ABMstateangle}(f) are measured in the C direction.
Furthermore, we utilized the theoretical curves of the ABM state to fit the data of the PCAR experiment of the Bi/Ni bilayer \cite{ZGH}. In Fig. \ref{fig:ABMstateangle}, the black and blue circles represent the experimental data \cite{ZGH}, while the red lines represent the theoretical fitting curves based on our theory. The normalized conductance spectra do not depend on spin polarization. Figures \ref{fig:ABMstateangle}(a) and \ref{fig:ABMstateangle}(b) show a single peak with values of $\varphi=0.25\pi$ and $\varphi=0.3\pi$, respectively; Figs. \ref{fig:ABMstateangle}(c) and \ref{fig:ABMstateangle}(d) show a double peak with a value of $\varphi=0.01\pi$ around zero; Figs. \ref{fig:ABMstateangle}(e) and \ref{fig:ABMstateangle}(f) show a single peak with values of $\varphi=0.29\pi$ and $\varphi=0.18\pi$, respectively. The angle parameter $\varphi$ mainly determines the peak features of the conductance spectra of the ABM state, including the number of the peaks. The angle parameter values shown in Fig. \ref{fig:ABMstateangle} are the best that we can find to fit these measured data and satisfy the features of the measured data. The value of $Z$ mainly determines the height of the conductance peak. The value of $\Gamma$ mainly changes the small details of the conductance spectra. We can easily find that the configuration of the ABM state shown in Figs. \ref{fig:pdtunnels}(c) and \ref{fig:pdtunnels}(d) satisfies the features of the measured conductance peaks in the three directions. Moreover, it is consistent with our theory that their experimental conductance spectra are independent of spin polarization. Therefore, the ABM state might be indicated in the bulk of the Bi layer of the Bi/Ni bilayer, as shown in Fig. \ref{fig:3Ddemonstrationmini}. We should consider the three-dimensional gap structure of the ABM state in the Bi/Ni bilayer, according to Figs. \ref{fig:3Ddemonstrationmini}, \ref{fig:pdtunnels}(c), \ref{fig:pdtunnels}(d) and \ref{fig:ABMstateangle}. Moreover, the axis of symmetry of the ABM state's gap is almost parallel to the B direction. It is enough that we only consider the 3D gap cross-section parallel to the Bi/Ni bilayer surface for the conductance spectra in the B direction, according to the symmetry of the ABM state's gap. However, we only obtain the AR in the particular cross-section of the 3D gap of the ABM state when the tip is in the A direction or the C direction. Thus we should consider an arbitrary cross-section of the 3D gap.

The gap values obtained by the fitting procedure range from 1.9 to 2.7 meV, as shown in Fig. \ref{fig:ABMstateangle}. The fitted gap values of the B direction are smaller than those of the A or C directions. We propose the inhomogeneity of the Bi/Ni bilayer samples should be the reason. The inhomogeneity of these samples may result from aging, air exposure, and imperfections during growth \cite{NodelessBiNiPhysRevLett.122.017002}.

\begin{figure}
\centering
\includegraphics[width=0.35\textwidth]{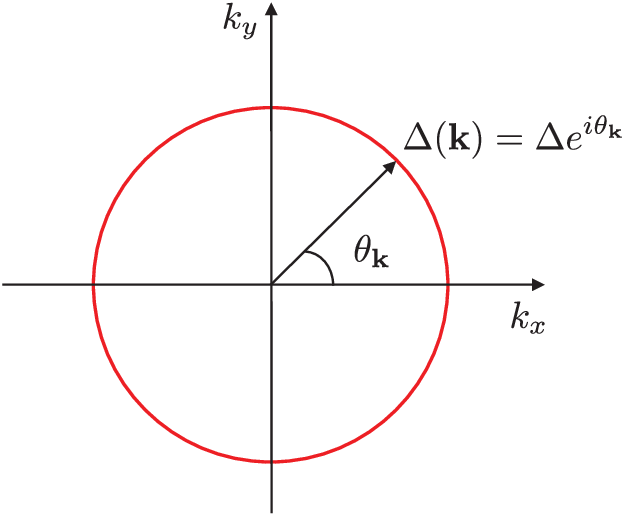}
\caption{\label{fig:chiralcrosssection}(Color online) The red circle represents the cross-section of the 3D gap of the ABM state in the case of $\omega=\pi/2$ and $\phi_{0}=\pi/2$, and this cross-section is the same as the chiral $p$-wave.}
\end{figure}

\section{Andreev reflection of the ABM state in an arbitrary cross-section of the gap}\label{sec:ABM_arbitrary_cross-section}
As said above, we should consider the gap structure of the ABM state in three dimensions for the PCAR experiments on the Bi/Ni bilayer. We have presented the conductance formula of an arbitrary cross-section of the 3D gap of the ABM state in Appendix \ref{Sec:arbcrosssectionABMformula}. As shown in Fig. \ref{fig:3Ddemonstrationmini}, when the tip is in the B direction, the wave vector $\mathbf{ k}_{B}$ of the incident electron can be decomposed into two components, $\mathbf{ k}_{\parallel}$ and $\mathbf{ k}_{\bot}$, which are parallel and perpendicular to the N-S interface, respectively. Therefore, these two wave vector components determine the cross-section of $\omega=0$ and $\phi_{0}=\mathrm{const.}$. Then the pair potential that quasiparticles experience in this cross-section is given by $\pm\Delta\sin\theta_{\mathbf{k}}e^{i\phi_{0}}\hat{\sigma}_{z}$, according to Eq. (\ref{eq:rotatedABMgap}). For example, the case of $\mathbf{ k}_{\parallel}$ parallel to the Bi/Ni bilayer surface corresponds to the cross-section of $\omega=0$ and $\phi_{0}=0$. As discussed above, the symmetrical axis of the 3D gap of the ABM state is almost along the B direction. The characteristics of the cross-sections of $\omega=0$ and $\phi_{0}=\mathrm{const.}$ are the same as that of the cross-section of $\omega=0$ and $\phi_{0}=0$, on account of the rotational symmetry of the 3D gap along the B direction.

As shown in Fig. \ref{fig:3Ddemonstrationmini}, for the case of measuring in the C direction, we can obtain the AR cross-section of $\omega=\pi/2$ and $\phi_{0}=\mathrm{const.}$ in the same way. When the $\mathbf{ k}_{\parallel}$ is parallel to the surface of the Bi/Ni bilayer, the corresponding cross-section is determined by $\omega=\pi/2$ and $\phi_{0}=0$, namely, $\Delta\cos\theta_{\mathbf{k}}\hat{\sigma}_{z}$, according to Eq. (\ref{eq:rotatedABMgap}). However, there is no rotational symmetry of the 3D gap along the C direction. Therefore, the characteristics of other cross-sections of the 3D gap in the C direction are different from that of the cross-section of $\omega=\pi/2$ and $\phi_{0}=0$. Particularly, when the $\mathbf{ k}_{\parallel}$ is perpendicular to the B direction, namely, the cross-section of $\omega=\pi/2$ and $\phi_{0}=\pi/2$, the corresponding cross-section of the 3D gap is given by $\Delta e^{\pm i\theta_{\mathbf{k}}}\hat{\sigma}_{z}$, according to Eq. (\ref{eq:rotatedABMgap}). As shown in Fig. \ref{fig:chiralcrosssection}, this cross-section is the same as the chiral $p$-wave. The AR spectra of the chiral $p$-wave are shown in Fig. \ref{fig:chiralABMstate}, and they are still of a single peak, similar to those of the cross-section of $\omega=\pi/2$ and $\phi_{0}=0$. However, they are isotropic. In the cross-section of $\omega=\pi/2$ and $\phi_{0}=\mathrm{const.}$, the corresponding conductance spectra are always of a single peak, and the height of the conductance peak decreases with the value of the parameter $\phi_{0}$ increasing from $0$ to $\pi/2$. The explanation of this behavior can be found in Appendix \ref{Sec:arbcrosssectionABMformula}. This important single peak characteristic guarantees that one will almost always obtain a single peak signal in the A or C direction in the experiment, as shown in Fig. \ref{fig:3Ddemonstrationmini}. For convenience, we utilize the cross-section of $\omega=\pi/2$ and $\phi_{0}=0$ to describe the main characteristics of AR in the C direction. The analysis of the A direction is identical to that of the C direction.

The chiral cross-section (namely, the cross-section of $\omega=\pi/2$ and $\phi_{0}=\pi/2$) is critical. We propose that the chiral cross-section of the ABM state might explain the broken time-reversal symmetry (TRS) determined by the polar Kerr effect measurements in the Bi/Ni bilayer \cite{Gonge1602579} and the time-domain THz spectroscopy (TDTS) where fully gapped superconductivity in the bulk of the system is proposed \cite{NodelessBiNiPhysRevLett.122.017002}. The Bi layer of the Bi/Ni bilayer has a finite thickness, and the PCAR experimental data \cite{ZGH} are obtained from three mutually perpendicular directions of the Bi/Ni bilayer. These indicate the ABM state's gap in the Bi/Ni bilayer should be understood from a three-dimensional perspective rather than a two-dimensional perspective. When the measurement is performed on the chiral cross-section of the ABM state, the characteristics of the chiral $p$-wave will be detected, including the broken TRS and fully gapped superconductivity. Here the chiral $p$-wave (i.e., $p+ip$) naturally has two $p$-wave components with equal magnitudes. This can explain well the TDTS of the Bi/Ni bilayer \cite{NodelessBiNiPhysRevLett.122.017002}, where the gap structure with approximately equal magnitudes for two $p$-wave components of the chiral $p$-wave is proposed. The characteristics of the chiral $p$-wave will manifest as long as the measurement is not perpendicular to the symmetrical axis of the 3D gap of the ABM state in Bi/Ni bilayers. Furthermore, due to the inhomogeneity of the Bi/Ni bilayer sample, the axes of symmetry of the ABM state's gap in different localities of a sample may only be roughly parallel to each other rather than completely parallel. Therefore, the characteristics of the chiral $p$-wave always manifest in various measurements.

\section{Discussion and conclusion}\label{Sec:DiscussionConclusion}
Our theoretical formalism by quantitatively describing the effects of spin polarization demonstrates that the PCAR experiments can be used to distinguish between the singlet pairing and the triplet pairing. The superconducting state of the Bi/Ni bilayers belongs to the triplet pairing rather than the singlet pairing since the PCAR spectroscopy results are independent of spin polarization. This critical feature can rule out the possibility of the $s$-wave and the $d$-wave.

As said above, the zero-bias conductance peak of the PCAR spectra originates from the $\pi$ phase difference in the cross-section of the gap. According to this simple physical picture, the zero-bias conductance peak signal of the Bi/Ni bilayer in the A and C directions indicates the existence of the $\pi$ phase difference in the cross-section of the gap. Note that the zero-bias conductance peak is robust for the Bi/Ni bilayers \cite{ZGH}. We should also emphasize that the values of the height of the normalized zero-bias conductance peak of the Bi/Ni bilayer are substantially higher than two, as shown in Figs. \ref{fig:ABMstateangle}(a), \ref{fig:ABMstateangle}(b), and \ref{fig:ABMstateangle}(e). This important feature also rules out the possibility of the $s$-wave since the corresponding values of the height of conductance within the gap ($eV\le \Delta$) are not more than two \cite{BTK,Daghero_2010}. In addition, the zero-bias conductance peak of the Bi/Ni bilayer does not originate from the effect of the thermal smearing on the normalized conductance, according to this feature.

We also rule out the Balian-Werthamer (BW) state of the $p$-wave because the anisotropic PCAR spectra of the Bi/Ni bilayer are not consistent with the isotropic conductance spectra of the BW state (see Appendix \ref{Sec:BWandchiralp}). We propose that the ABM state of the triplet $p$-wave seizes the essential features of the superconducting state of the Bi/Ni bilayers, according to our theoretical fitting curves. The mechanism of the ABM state in the Bi/Ni bilayers might be complex. The external perturbation to this system during probing might slightly change the gap values of the ABM state. The qualitative features of the ABM state in the Bi/Ni bilayers are robust, which makes the fitting curves satisfy the PCAR data. We think the ABM state is a simple and elegant description of the superconducting state in the Bi/Ni bilayers.

The features of the PCAR data of the Bi/Ni bilayers indicate the unconventional pairing in this system. However, to provide more evidence in favor of the ABM state, other experimental works in the Bi/Ni bilayers are expected in the future, including phase-sensitive symmetry tests and non-phase-sensitive techniques. To further ensure the ABM state in the Bi/Ni bilayers is meaningful. This material will become the first discovered solid material where the ABM state exists if more experiments confirm this.

In conclusion, we utilize the theoretical formalism to demonstrate that the AR of triplet pairing superconductivity is independent of spin polarization, contrary to the singlet pairing case. Our theoretical conductance spectra of the ABM state can explain the main features of the PCAR spectroscopy results of the epitaxial Bi/Ni bilayers. In addition, the candidate ABM state demonstrates the exact conductance formula for an arbitrary cross-section of the 3D gap.

\begin{acknowledgments}
We are grateful for the invaluable discussions with T. Y. Chen, C. L. Chien, X. F. Jin, and T. K. Lee. This work was supported by the National Key Research and Development Program of China Grant No. 2022YFA1404204, the National Natural Science Foundation of China (Grants No. 11625416 and No. 12274086).
\end{acknowledgments}

\appendix
\section{The unitary solution and the construction of the wave function at the superconductor side}\label{sec:appendixunitarysolution}
We can utilize the four-component notations, $ \mathbf{a_{k}}=(a_{\mathbf{k}\uparrow},a_{\mathbf{k}\downarrow},a_{-\mathbf{k}\uparrow}^{\dagger},a_{-\mathbf{k}\downarrow}^{\dagger})^{T}$ and $\mathbf{\alpha_{k}}=(\alpha_{\mathbf{k}\uparrow},\alpha_{\mathbf{k}\downarrow},\alpha_{-\mathbf{k}\uparrow}^{\dagger},\alpha_{-\mathbf{k}\downarrow}^{\dagger})^{T}$, to make Eq. (\ref{eq:quasiparticle}) compact: $\mathbf{a_{k}}= U_{\mathbf{k}}\mathbf{\alpha_{k}}$ with
\begin{equation}\label{eq:unitaritycondition}
U_{\mathbf{k}}=\left( \begin{array}{cc}
\hat{u}_{\mathbf{k}}&\hat{v}_{\mathbf{k}}\\
\hat{v}^{*}_{-\mathbf{k}}&\hat{u}^{*}_{-\mathbf{k}}
\end{array} \right) ~\textrm{and} ~U_{\mathbf{k}}U_{\mathbf{k}}^{\dagger}=1,
\end{equation}
where $\hat{u}_{\mathbf{k}}$ and $\hat{v}_{\mathbf{k}}$ are $2\times2$ matrices, i.e.,
\begin{equation}\label{eq:uvmatrices}
\hat{u}_{\mathbf{k}}=\left( \begin{array}{cc}
u_{\mathbf{k}\uparrow\uparrow}&u_{\mathbf{k}\uparrow\downarrow}\\
u_{\mathbf{k}\downarrow\uparrow}&u_{\mathbf{k}\downarrow\downarrow}
\end{array} \right) ~\textrm{and} ~\hat{v}_{\mathbf{k}}=\left( \begin{array}{cc}
v_{\mathbf{k}\uparrow\uparrow}&v_{\mathbf{k}\uparrow\downarrow}\\
v_{\mathbf{k}\downarrow\uparrow}&v_{\mathbf{k}\downarrow\downarrow}
\end{array} \right).
\end{equation}
According to Eq. (\ref{eq:unitaritycondition}), we can obtain
\begin{equation}\label{eq:alfarelation}
\mathbf{\alpha_{k}}= U_{\mathbf{k}}^{\dagger}\mathbf{a_{k}}.
\end{equation}
By Eq. (\ref{eq:alfarelation}), we can construct Eq. (\ref{eq:superstate}). $\psi_{S}(\mathbf{ r})$ is combined by the transmitted quasiparticle of the same side of the Fermi surface (i.e., the electron-like quasiparticle) and the transmitted quasiparticle crossing through the Fermi surface (i.e., the hole-like quasiparticle) \cite{BTK}. The electron-like quasiparticle wave function is $e^{i\mathbf{ k}_{\parallel}\cdot \mathbf{ r}_{\parallel}}e^{iq_{+}x}(
u^{(+)}_{\mathbf{k}\uparrow\uparrow},u^{(+)}_{\mathbf{k}\downarrow\uparrow},v^{*(+)}_{\mathbf{-k}\uparrow\uparrow},v^{*(+)}_{\mathbf{-k}\downarrow\uparrow})^{T}$ which corresponds to the quasiparticle's creation operator $\alpha^{\dagger}_{\mathbf{ k}\uparrow}$ \cite{P.G.DeGennes}. And we have the relation $\alpha_{\mathbf{k}\uparrow}=u_{\mathbf{k}\uparrow\uparrow}^{*}a_{\mathbf{k}\uparrow}+u_{\mathbf{k}\downarrow\uparrow}^{*}a_{\mathbf{k}\downarrow}+v_{-\mathbf{k}\uparrow\uparrow}a_{-\mathbf{k}\uparrow}^{\dagger}+v_{-\mathbf{k}\downarrow\uparrow}a_{-\mathbf{k}\downarrow}^{\dagger}$. It is the same way to obtain the hole-like quasiparticle wave function. Thus, we can construct Eq. (\ref{eq:superstate}) using the elements of the matrix $U_{\mathbf{k}}$. The wave function components corresponding to other quasiparticle $\alpha_{\mathbf{k}\downarrow}$ can be neglected in the case of the unitary $\hat{\Delta}(\mathbf{k})$, as explained below. For the unitary solution \cite{Sigrist}
\begin{flalign}
&\hat{u}_{\mathbf{k}}=\frac{[E_{\mathbf{k}}+\epsilon(\mathbf{ k})]\hat{\sigma}_{0}}
{\left\{[E_{\mathbf{k}}+\epsilon(\mathbf{ k})]^{2}+\frac{1}{2}\mathbf{tr}\hat{\Delta}\hat{\Delta}^{\dagger}(\mathbf{ k})\right\}^{1/2}},
                                                \nonumber\\
&\hat{v}_{\mathbf{k}}=\frac{-\hat{\Delta}(\mathbf{ k})}
{\left\{[E_{\mathbf{k}}+\epsilon(\mathbf{ k})]^{2}+\frac{1}{2}\mathbf{tr}\hat{\Delta}\hat{\Delta}^{\dagger}(\mathbf{ k})\right\}^{1/2}},
\label{eq:unitary}
\end{flalign}
where $E_{\mathbf{k}}=\left[\epsilon^{2}(\mathbf{ k})+\frac{1}{2}\mathbf{tr}\hat{\Delta}\hat{\Delta}^{\dagger}(\mathbf{ k})\right]^{1/2}$ is the energy spectrum of the superconducting quasiparticles, and the gap function $\hat{\Delta}(\mathbf{ k})$ is a matrix. For the singlet pairing case
\begin{equation}\label{eq:singletpairing}
\hat{\Delta}(\mathbf{ k})=i\hat{\sigma}_{y}\psi(\mathbf{ k})=\left(\begin{array}{cc}
0&\psi(\mathbf{ k})\\
-\psi(\mathbf{ k})&0
\end{array}\right),
\end{equation}
where $\psi(\mathbf{ k})$ is an even function. For the triplet pairing case
\begin{flalign}
\hat{\Delta}(\mathbf{ k})&=i(\mathbf{ d}(\mathbf{ k})\cdot\hat{\boldsymbol{\sigma}})\hat{\sigma}_{y}
                                                             \nonumber\\
&=\left(\begin{array}{cc}-d_{x}(\mathbf{ k})+id_{y}(\mathbf{ k})&d_{z}(\mathbf{ k})\\d_{z}(\mathbf{ k})&d_{x}(\mathbf{ k})+id_{y}(\mathbf{ k})
\end{array}\right),
\label{eq:tripletpairing}
\end{flalign}
where $\mathbf{ d}(\mathbf{ k})$ is an odd vectorial function. Since $\alpha_{\mathbf{k}\downarrow}=u_{\mathbf{k}\uparrow\downarrow}^{*}a_{\mathbf{k}\uparrow}+u_{\mathbf{k}\downarrow\downarrow}^{*}a_{\mathbf{k}\downarrow}+v_{-\mathbf{k}\uparrow\downarrow}a_{-\mathbf{k}\uparrow}^{\dagger}+v_{-\mathbf{k}\downarrow\downarrow}a_{-\mathbf{k}\downarrow}^{\dagger}$, we have $u_{\mathbf{k}\uparrow\downarrow}^{*}=0$ and thus $\alpha_{\mathbf{k}\downarrow}=u_{\mathbf{k}\downarrow\downarrow}^{*}a_{\mathbf{k}\downarrow}+v_{-\mathbf{k}\uparrow\downarrow}a_{-\mathbf{k}\uparrow}^{\dagger}+v_{-\mathbf{k}\downarrow\downarrow}a_{-\mathbf{k}\downarrow}^{\dagger}$ for the unitary solution of superconductivity according to Eq. (\ref{eq:unitary}). We know from Eq. (\ref{eq:normalstate}) that the spin of the incident electron wave is up. Thus the spin of the transmitted electron is also up, i.e., the space of $\alpha^{\dagger}_{\mathbf{k}\downarrow}$ is orthogonal to the incident electron. Then the transmitted electron with spin up in the superconductor will induce the quasiparticle corresponding to the annihilation operator $\alpha_{\mathbf{k}\uparrow}=u_{\mathbf{k}\uparrow\uparrow}^{*}a_{\mathbf{k}\uparrow}+v_{-\mathbf{k}\uparrow\uparrow}a_{-\mathbf{k}\uparrow}^{\dagger}+v_{-\mathbf{k}\downarrow\uparrow}a_{-\mathbf{k}\downarrow}^{\dagger}$ by acting as the component corresponding to $a_{\mathbf{k}\uparrow}$ of the quasiparticle $\alpha_{\mathbf{k}\uparrow}$. However, because there is no transmitted electron with spin down in the superconductor as the inducible factor, the transmitted electron with spin up is unable to induce the other quasiparticle corresponding to the annihilation operator $\alpha_{\mathbf{k}\downarrow}=u_{\mathbf{k}\downarrow\downarrow}^{*}a_{\mathbf{k}\downarrow}+v_{-\mathbf{k}\uparrow\downarrow}a_{-\mathbf{k}\uparrow}^{\dagger}+v_{-\mathbf{k}\downarrow\downarrow}a_{-\mathbf{k}\downarrow}^{\dagger}$. As a result, the wave function components associated with the quasiparticle $\alpha_{\mathbf{k}\downarrow}$ can be neglected.

\section{The properties and the calculated coefficients of the ABM state}\label{sec:propertiesofABMorder}
The $\mathbf{ d}(\mathbf{ k})$ corresponding to the form of the ABM state's order parameter (i.e., Eq. (\ref{eq:ABMorderparameter})) is given by $\mathbf{ d}(\mathbf{ k})=\Delta(e^{i\phi_{\mathbf{k}}}\sin\theta_{\mathbf{k}},0,0)=\Delta(\hat{k}_{x}+i\hat{k}_{y},0,0)$, where $\hat{k}_{x}$ ($\hat{k}_{y}$) is a linear combination of angular-momentum eigenstates $Y_{1}^{m}(\hat{\mathbf{k}})$ ($m=0,\pm 1$), according to Eq. (\ref{eq:tripletpairing}). The relation $\hat{k}_{x}+i\hat{k}_{y}\propto Y_{1}^{+1}$ indicates a finite orbital-angular-momentum projection along the $\hat{z}$-axis parallel to the axis of symmetry of the ABM state's gap \cite{Vollhardt}. The ABM state breaks time-reversal symmetry, i.e., $K\hat{\Delta}(\mathbf{ k})\ne \hat{\Delta}(\mathbf{ k})e^{i\Phi}$, where $K$ is the time-reversal operator and $\Phi$ is a phase factor related to the U(1) gauge transformation \cite{Sigrist}, according to the relation $K\mathbf{ d}(\mathbf{ k})=-\mathbf{ d}^{*}(\mathbf{ -k})$.

The calculated $a_{1}$, $a_{2}$, $b$, $c$, and $d$ in the case of the cross-section of $\phi_{0}=0$ are listed in Table \ref{tab:ABMresult}. Moreover, Table \ref{tab:3DABMresult} gives the results of the 3D case.

\begin{table*}
\caption{\label{tab:ABMresult}
Here\quad$\hat{u}_{\mathbf{q}}^{(\pm)}=\sqrt{\frac{1}{2}\left\lgroup1\pm\epsilon^{\pm}(\mathbf{q})/\left(|E_{\mathbf{q}}|+i\Gamma\right)\right\rgroup}\hat{\sigma}_{0}$,
$\hat{v}_{\mathbf{q}}^{(\pm)}=\rm{sgn}(\sin\theta^{\pm}_{\mathbf{q}})\sqrt{\frac{1}{2}\left\lgroup1\mp\epsilon^{\pm}( \mathbf{q} )/\left(|E_{\mathbf{q}}|+i\Gamma\right)\right\rgroup}\hat{\sigma}_{z}$,
$p=(Z_{\mathbf{q}}^{2}+1)u_{\mathbf{q}\uparrow\uparrow}^{(+)}v_{\mathbf{-q}\uparrow\uparrow}^{*(-)}-Z_{\mathbf{q}}^{2}u_{\mathbf{q}\uparrow\uparrow}^{(-)}v_{\mathbf{-q}\uparrow\uparrow}^{*(+)}$,
\quad$\epsilon^{\pm}(\mathbf{q})=\sqrt{(|E_{\mathbf{q}}|+i\Gamma)^{2}-\Delta^{2}(\theta^{\pm}_{\mathbf{q}})}$,
 \quad$\Delta(\theta^{\pm}_{\mathbf{q}})=\Delta\sin(\theta^{\pm}_{\mathbf{q}})$,\quad$\theta^{+}_{\mathbf{q}}=\theta_{\mathbf{q}}-\varphi$,\quad$\theta^{-}_{\mathbf{q}}=\pi-\theta_{\mathbf{q}}-\varphi$,\quad$Z_{\mathbf{q}}=Z/\cos\theta_{N}^{\mathbf{q}}$,\quad the incident angle of the electron given by $\theta_{N}^{\mathbf{q}}=\theta_{\mathbf{q}}$, \quad$Z=mU/\hbar^{2}k_{F}$, and $\theta_{\mathbf{q}}\in\left[0,2\pi\right]$.}
\begin{ruledtabular}
\begin{tabular}{cccccc}
&$a_{1}$&$a_{2}$&b&c&d\\
\hline
\\
ABM state (2D case)&$0$&$\frac{v_{\mathbf{-q}\uparrow\uparrow}^{*(+)}v_{\mathbf{-q}\uparrow\uparrow}^{*(-)}}{p}$
&$\frac{Z_{\mathbf{q}}(Z_{\mathbf{q}}+i)(u_{\mathbf{q}\uparrow\uparrow}^{(-)}v_{\mathbf{-q}\uparrow\uparrow}^{*(+)}-u_{\mathbf{q}\uparrow\uparrow}^{(+)}v_{\mathbf{-q}\uparrow\uparrow}^{*(-)})}{p}$
&$\frac{-i(Z_{\mathbf{q}}+i)v_{\mathbf{-q}\uparrow\uparrow}^{*(-)}}{p}$
&$\frac{iZ_{\mathbf{q}}v_{\mathbf{-q}\uparrow\uparrow}^{*(+)}}{p}$\\
\\
\end{tabular}
\end{ruledtabular}
\end{table*}

\begin{table*}
\caption{\label{tab:3DABMresult}
Here\quad$u_{\mathbf{k}\uparrow\uparrow}^{(\pm)}=\sqrt{\frac{1}{2}\left\lgroup1\pm\epsilon^{\pm}(\mathbf{k})/\left(|E_{\mathbf{k}}|+i\Gamma\right)\right\rgroup}$,
$v_{\mathbf{-k}\uparrow\uparrow}^{*(\pm)}=\Delta^{*}_{I}(\theta_{\mathbf{k}}^{\pm},\phi_{\mathbf{k}},\omega)/\sqrt{2\left(|E_{\mathbf{k}}|+i\Gamma\right)\left\lgroup\left(|E_{\mathbf{k}}|+i\Gamma\right)\pm\epsilon^{\pm}( \mathbf{k} )\right\rgroup}$,
$p=(Z_{\mathbf{k}}^{2}+1)u_{\mathbf{k}\uparrow\uparrow}^{(+)}v_{\mathbf{-k}\uparrow\uparrow}^{*(-)}-Z_{\mathbf{k}}^{2}u_{\mathbf{k}\uparrow\uparrow}^{(-)}v_{\mathbf{-k}\uparrow\uparrow}^{*(+)}$,
\quad$\epsilon^{\pm}(\mathbf{k})=\sqrt{\left(|E_{\mathbf{k}}|+i\Gamma\right)^{2}-|\Delta_{I}(\theta^{\pm}_{\mathbf{k}},\phi_{\mathbf{k}},\omega)|^{2}}$,
 \quad$\Delta_{I}(\theta_{\mathbf{k}},\phi_{\mathbf{k}},\omega)=-\Delta(\sin\theta_{\mathbf{k}}\cos\phi_{\mathbf{k}}\cos\omega-\cos\theta_{\mathbf{k}}\sin\omega+i\sin\theta_{\mathbf{k}}\sin\phi_{\mathbf{k}})$,\quad$\theta_{\mathbf{k}}^{+}=\theta_{\mathbf{k}}$,\quad$\theta_{\mathbf{k}}^{-}=\pi-\theta_{\mathbf{k}}$,\quad$Z_{\mathbf{k}}=Z/\cos\theta_{N}^{\mathbf{k}}$,\quad$\theta_{N}^{\mathbf{k}}=\theta_{\mathbf{k}}$, and $Z=mU/\hbar^{2}k_{F}$.}
\begin{ruledtabular}
\begin{tabular}{cccccc}
&$a_{1}$&$a_{2}$&b&c&d\\
\hline
\\
ABM state (3D case)&$0$&$\frac{v_{\mathbf{-k}\uparrow\uparrow}^{*(+)}v_{\mathbf{-k}\uparrow\uparrow}^{*(-)}}{p}$
&$\frac{Z_{\mathbf{k}}(Z_{\mathbf{k}}+i)(u_{\mathbf{k}\uparrow\uparrow}^{(-)}v_{\mathbf{-k}\uparrow\uparrow}^{*(+)}-u_{\mathbf{k}\uparrow\uparrow}^{(+)}v_{\mathbf{-k}\uparrow\uparrow}^{*(-)})}{p}$
&$\frac{-i(Z_{\mathbf{k}}+i)v_{\mathbf{-k}\uparrow\uparrow}^{*(-)}}{p}$
&$\frac{iZ_{\mathbf{k}}v_{\mathbf{-k}\uparrow\uparrow}^{*(+)}}{p}$\\
\\
\end{tabular}
\end{ruledtabular}
\end{table*}

\section{The conductance formula for Andreev reflection of the ABM state in an arbitrary cross section}\label{Sec:arbcrosssectionABMformula}
Now let us demonstrate how to get the exact conductance formula of an arbitrary cross-section of the 3D gap by the example of the ABM state. We need some tricks to link the gap equation to the wave vector direction of the incident electron from the N side since the incident electron with different wave vector directions will experience different phases of the pair potential at the S side. The wave vector direction of an incident electron is given by $\hat{\mathbf{k}}\equiv(\hat{k}_{x},\hat{k}_{y},\hat{k}_{z})\equiv(\sin\theta_{\mathbf{k}}\cos\phi_{\mathbf{k}},\sin\theta_{\mathbf{k}}\sin\phi_{\mathbf{k}},\cos\theta_{\mathbf{k}})$ expressed in the coordinate system $I$ (the definition of the coordinate system $I$ can be found in Sec. \ref{Sec:conductanceABM2D}, and $\mathbf{k}$ is the wave vector of the incident electron). We still restrict the axis of symmetry of the ABM state's gap to the $I_{x}-I_{z}$ plane and use the parameter $\omega$ defined in Sec. \ref{Sec:conductanceABM2D}. Then rotate the coordinate system $I$ around the $I_{y}$-axis to make the $I_{z}$-axis coincide with the axis of symmetry of the ABM state's gap, and, in the new coordinate system $I^{\prime}$, the wave vector direction $\hat{\mathbf{k}}$ is rewritten as $\hat{\mathbf{k}}^{\prime T}\equiv R(\hat{I}_{y},-\omega)\hat{\mathbf{k}}^{T}$. Here $R(\hat{I}_{y},-\omega)$ is given by
\begin{equation}\label{eq:rotationABM}
 R(\hat{I}_{y},-\omega)=\left(\begin{array}{ccc}\cos\omega&0&-\sin\omega\\0&1&0\\ \sin\omega&0&\cos\omega
\end{array}\right).
\end{equation}
Then we obtain
\begin{flalign}
\hat{\mathbf{k}}^{\prime}=&(\sin\theta_{\mathbf{k}}\cos\phi_{\mathbf{k}}\cos\omega-\cos\theta_{\mathbf{k}}\sin\omega,
                                                       \nonumber\\
&\sin\theta_{\mathbf{k}}\sin\phi_{\mathbf{k}}, \sin\theta_{\mathbf{k}}\cos\phi_{\mathbf{k}}\sin\omega+\cos\theta_{\mathbf{k}}\cos\omega)
\label{eq:newcoordinate}.
\end{flalign}
Thus the gap equation of the ABM state can be rewritten as
\begin{flalign}
&\hat{\Delta}(\mathbf{ k})=-\Delta(\hat{k}_{x}^{\prime}+i\hat{k}_{y}^{\prime})\hat{\sigma}_{z}=
-\Delta(\sin\theta_{\mathbf{k}}\cos\phi_{\mathbf{k}}\cos\omega
                                                \nonumber\\
&-\cos\theta_{\mathbf{k}}\sin\omega+i\sin\theta_{\mathbf{k}}\sin\phi_{\mathbf{k}})\hat{\sigma}_{z}.
\label{eq:rotatedABMgap}
\end{flalign}
Now, this gap function is linked to the direction of the wave vector of the incident electron. Therefore, the expressions of $\hat{u}_{\mathbf{k}}$ and $\hat{v}_{\mathbf{k}}$ linked to the wave vector of the incident electron are given by
\begin{flalign}
&\hat{u}_{\mathbf{k}}=\sqrt{\frac{1}{2}\left(1+\epsilon(\mathbf{ k})/E_{\mathbf{k}}\right)}\hat{\sigma}_{0},
                                              \nonumber\\
&\hat{v}_{\mathbf{k}}=\frac{-\hat{\Delta}(\mathbf{ k})}{\sqrt{2E_{\mathbf{k}}(E_{\mathbf{k}}+\epsilon(\mathbf{ k}))}}.
\label{eq:3Dukvkexpression}
\end{flalign}
The corresponding calculated coefficients $a_{1}$, $a_{2}$, $b$, $c$, and $d$ for the incident electron with the wave vector $\mathbf{k}$ are listed in Table \ref{tab:3DABMresult}. The expressions of these coefficients are almost the same as that of the 2D case. Thus we can obtain the corresponding conductance formula by inserting these coefficients into Eq. (\ref{eq:generalconductance}).

The parameters $\omega$ and $\phi_{0}$ (the definition of $\phi_{0}$ can be found in Sec. \ref{Sec:conductanceABM2D}) will determine the cross-section of the gap of the ABM state, in which the AR process occurs. Moreover, the AR cross-section will be determined by the configuration of connecting the tip and the sample of the ABM state in the PCAR experiment.

\begin{figure}
\centering
\includegraphics[width=0.45\textwidth]{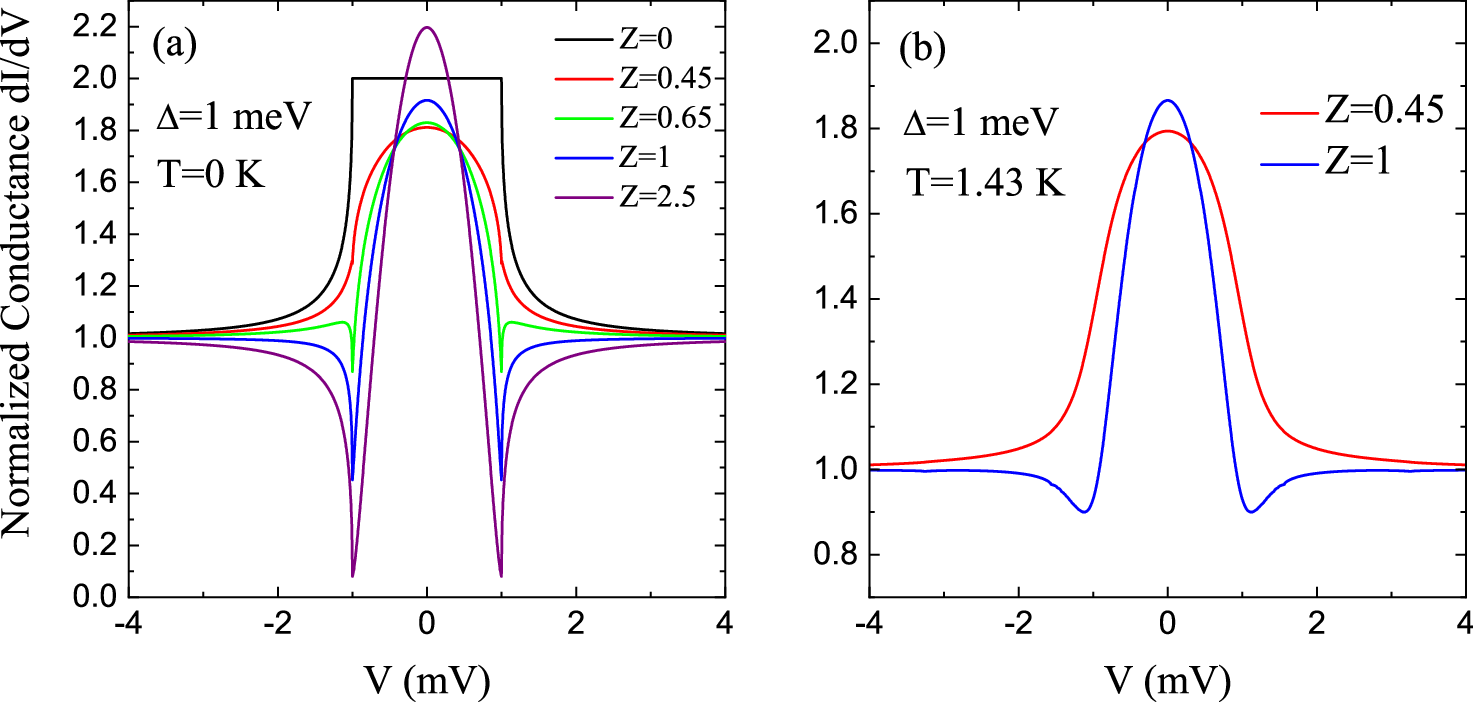}
\caption{\label{fig:chiralABMstate}(Color online) The normalized conductance spectra of the BW state in the case of the gap cross-section parallel to the $x$-$y$ plane and the chiral $p$-wave state.}
\end{figure}

The orbital part of the ABM state in the cross-section of $\omega=\pi/2$ and $\phi_{0}=\pi/2$ can be rewritten as $e^{\pm i\theta_{\mathbf{k}}}=\cos\theta_{\mathbf{k}}\pm i\sin\theta_{\mathbf{k}}$. The first component $\cos\theta_{\mathbf{k}}$ yields the zero-bias conductance peak, which is the same as that of the cross-section of $\omega=\pi/2$ and $\phi_{0}=0$. The second component $\pm i\sin\theta_{\mathbf{k}}$ is the same as the cross-section of $\omega=0$ and $\phi_{0}=0$. Both of these two components contribute to the conductance of the ABM state in the cross-section of $\omega=\pi/2$ and $\phi_{0}=\pi/2$, which is a linear combination of the cases indicated by Figs. \ref{fig:pdtunnels}(c) and \ref{fig:pdtunnels}(d). In the cross-section of $\omega=\pi/2$ and $\phi_{0}=\mathrm{const.}$, the corresponding orbital part is given by $\cos\theta_{\mathbf{k}}\pm i\sin\phi_{0}\sin\theta_{\mathbf{k}}$. Thus this case is also a linear combination of the cases indicated by Figs. \ref{fig:pdtunnels}(c) and \ref{fig:pdtunnels}(d). As the value of the parameter $\phi_{0}$ increases from $0$ to $\pi/2$, the weight factor $\sin\phi_{0}$ of the second component increases as well, and thus the height of the zero-bias conductance peak decreases. In addition, the transmission of electrons becomes weaker and more directional around the normal to the N-S interface as the value of $Z$ increases, according to $\frac{1}{(Z/\cos\theta_{N}^{\mathbf{k}})^{2}+1}$ (i.e., the transparency of an N-S interface). Thus as the value of $Z$ increases, the zero-bias peak becomes sharper due to the more contribution from the component $\cos\theta_{\mathbf{k}}$. In the chiral cross-section case, the zero-bias peak is sharp enough when $Z=0.35$, which is around the minimum value of $Z$ obtained from our fits on the PCAR spectra of Bi/Ni bilayers, as indicated in Fig. \ref{fig:chiralABMstate}.

\begin{figure}
\centering
\includegraphics[width=0.45\textwidth]{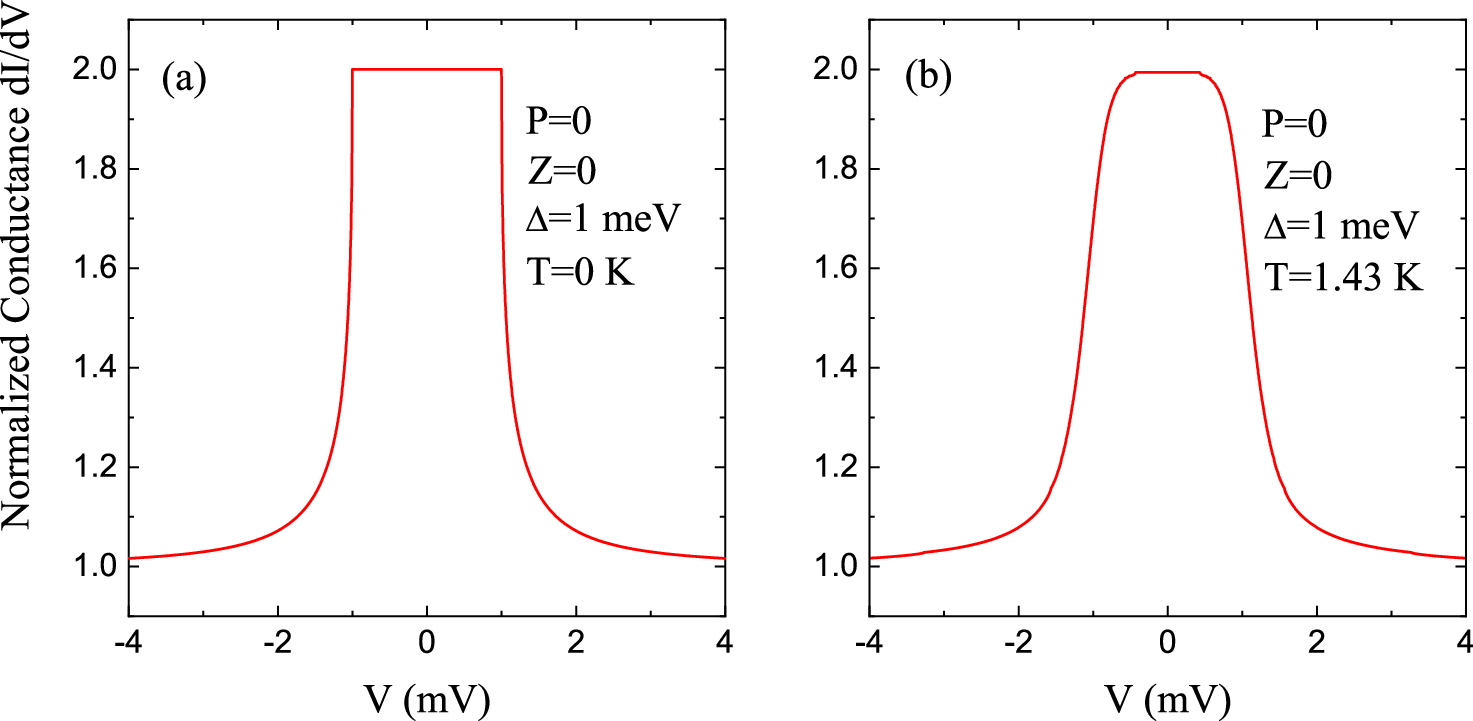}
\caption{\label{fig:BWstate}(Color online) The normalized conductance spectra of the BW state in the case of $Z=\alpha=0$.}
\end{figure}

\section{Conductance spectra for the BW state and the chiral p wave state}\label{Sec:BWandchiralp}
The BW state is the second important state of the $p$-wave pairing \cite{Balian,Leggett}. Its odd vectorial function is given by $\mathbf{ d}(\mathbf{ k})=\Delta\hat{\mathbf{k}}$, and the corresponding order parameter is given by
\begin{flalign}\label{eq:BWorderparameter}
\hat{\Delta}(\mathbf{ k})=\Delta\left(\begin{array}{cc}
-\hat{k}_{x}+i\hat{k}_{y}&\hat{k}_{z}\\
\hat{k}_{z}&\hat{k}_{x}+i\hat{k}_{y}
\end{array}\right),
\end{flalign}
where $\hat{k}_{x}=\sin\theta_{\mathbf{k}}\cos\phi_{\mathbf{k}}$, $\hat{k}_{y}=\sin\theta_{\mathbf{k}}\sin\phi_{\mathbf{k}}$ and $\hat{k}_{z}=\cos\theta_{\mathbf{k}}$. The BW state is unitary and its time-reversal symmetry remains unbroken \cite{Sigrist,Mackenzie,Honerkamp:1998wh}. The corresponding AR in an arbitrary cross-section of the gap of the BW state is restricted to the condition:
\begin{flalign}\label{eq:BWcondition}
&(Z_{\mathbf{k}}-i)u_{\mathbf{k}\uparrow\uparrow}^{(+)}\left[-Z_{\mathbf{k}}(2Z_{\mathbf{k}}+2i+\alpha)v_{-\mathbf{k}\uparrow\uparrow}^{*(+)}v_{-\mathbf{k}\downarrow\uparrow}^{*(-)}\right.
                       \nonumber\\
&\left.+(2Z_{\mathbf{k}}+\alpha)(Z_{\mathbf{k}}+i)v_{-\mathbf{k}\uparrow\uparrow}^{*(-)}v_{-\mathbf{k}\downarrow\uparrow}^{*(+)}\right]=0.
\end{flalign}
The corresponding AR coefficients are given by
\begin{flalign}\label{eq:BWARgeneralcoefficients}
&a_{1}=[-i(Z_{\mathbf{k}}+i)v_{-\mathbf{k}\downarrow\uparrow}^{*(+)}v_{-\mathbf{k}\uparrow\uparrow}^{*(-)}+iZ_{\mathbf{k}}v_{-\mathbf{k}\downarrow\uparrow}^{*(-)}v_{-\mathbf{k}\uparrow\uparrow}^{*(+)}]/p,
               \nonumber\\
&a_{2}=v_{-\mathbf{k}\uparrow\uparrow}^{*(+)}v_{-\mathbf{k}\uparrow\uparrow}^{*(-)}/p,
                \nonumber\\
&b=-Z_{\mathbf{k}}(Z_{\mathbf{k}}+i)(u_{\mathbf{k}\uparrow\uparrow}^{(+)}v_{-\mathbf{k}\uparrow\uparrow}^{*(-)}-u_{\mathbf{k}\uparrow\uparrow}^{(-)}v_{-\mathbf{k}\uparrow\uparrow}^{*(+)})/p,
                   \nonumber\\
&c=-i(Z_{\mathbf{k}}+i)v_{-\mathbf{k}\uparrow\uparrow}^{*(-)}/p,
                      \nonumber\\
&d=iZ_{\mathbf{k}}v_{-\mathbf{k}\uparrow\uparrow}^{*(+)}/p,
\end{flalign}
with $p=(Z_{\mathbf{k}}^{2}+1)u_{\mathbf{k}\uparrow\uparrow}^{(+)}v_{-\mathbf{k}\uparrow\uparrow}^{*(-)}-Z_{\mathbf{k}}^{2}u_{\mathbf{k}\uparrow\uparrow}^{(-)}v_{-\mathbf{k}\uparrow\uparrow}^{*(+)}$ and $Z_{\mathbf{k}}=Z/\cos\theta_{N}^{\mathbf{k}}$. Here $\theta_{N}^{\mathbf{k}}$ is the incident angle of the electron. Only two cases satisfy Eq. (\ref{eq:BWcondition}). The first case is when $Z=\alpha=0$. Then, $b=d=0$, $a_{1}=v_{-\mathbf{k}\downarrow\uparrow}^{*(+)}/u_{\mathbf{k}\uparrow\uparrow}^{(+)}$ and $a_{2}=v_{-\mathbf{k}\uparrow\uparrow}^{*(+)}/u_{\mathbf{k}\uparrow\uparrow}^{(+)}$. In this case, we can easily obtain
\begin{equation}\label{eq:geforBW}
g(E)=1+\left|\frac{E-\sqrt{E^{2}-\Delta^{2}}}{E+\sqrt{E^{2}-\Delta^{2}}}\right|.
\end{equation}
We can deduce from Eq. (\ref{eq:generalconductance}) that the corresponding normalized conductance is isotropic. Therefore, its conductance spectra are not related to the angle parameter $\varphi$, as shown in Fig. \ref{fig:BWstate}. Since Eq. (\ref{eq:geforBW}) is the same as that of the $s$-wave at $Z=0$, the conductance spectra are the same as those of the $s$-wave at $Z=0$, as shown in Fig. \ref{fig:BWstate}. The second case satisfying Eq. (\ref{eq:BWcondition}) is when the cross-section of the gap of the BW state is parallel to the $x$-$y$ plane, i.e., $\mathbf{ d}(\mathbf{ k})=\Delta(\hat{k}_{x},\hat{k}_{y},0)$ with $\theta_{\mathbf{k}}=\pi/2$. In this case, its order parameter is given by
\begin{flalign}\label{eq:BWxyorderparameter}
\hat{\Delta}(\mathbf{ k})=\Delta\left(\begin{array}{cc}
-e^{-i\phi_{\mathbf{k}}}&0\\
0&e^{i\phi_{\mathbf{k}}}
\end{array}\right),
\end{flalign}
which indicates $v_{-\mathbf{k}\downarrow\uparrow}^{*(\pm)}=0$ and $a_{1}=0$, according to Eqs. (\ref{eq:unitary}) and (\ref{eq:BWARgeneralcoefficients}). The corresponding conductance spectra are isotropic and of a single peak, and do not depend on spin polarization, as shown in Fig. \ref{fig:chiralABMstate}.

The order parameter of the $p+ip$ state (in two dimensions) is given by
\begin{equation}\label{eq:chirald}
\hat{\Delta}(\mathbf{ q})=-\Delta e^{i\phi_{\mathbf{q}}}\hat{\sigma}_{z},
\end{equation}
where $\phi_{\mathbf{q}}\in\left[0,2\pi\right]$. The chiral $p$-wave belongs to the unitary solution, according to $\hat{\Delta}(\mathbf{ q})\hat{\Delta}^{\dagger}(\mathbf{ q})=\Delta^{2}\hat{\sigma}_{0}$. Thus we can utilize Eq. (\ref{eq:unitary}) to obtain its coherence factors, which are elements of matrixes
\begin{flalign}
&\hat{u}_{\mathbf{q}}=\sqrt{\frac{1}{2}\left(1+\epsilon(\mathbf{ q})/E_{\mathbf{q}}\right)}\hat{\sigma}_{0},
                                                \nonumber\\
&\hat{v}_{\mathbf{q}}=\frac{-\hat{\Delta}(\mathbf{ q})}{\sqrt{2E_{\mathbf{q}}(E_{\mathbf{q}}+\epsilon(\mathbf{ q}))}},
\label{eq:chiraluv}
\end{flalign}
where $E_{\mathbf{q}}=\sqrt{\epsilon^{2}(\mathbf{ q})+\Delta^{2}}$. Following the same procedure, we can quickly obtain the coefficients $a_{1}$, $a_{2}$, $b$, $c$, and $d$, which are essentially the same as the ABM state case. Its conductance is also not related to spin polarization. The conductance spectra of the chiral $p$-wave state are identical to that of the BW state in the case of the gap cross-section parallel to the $x$-$y$ plane \cite{Honerkamp:1998wh}, as shown in Fig. \ref{fig:chiralABMstate}.

%\bibliography{ARS}

\end{document}